\documentclass[a4paper]{aa}
\usepackage[T1]{fontenc}
\usepackage[utf8]{inputenc}
\usepackage{graphicx}
\usepackage{color}
\usepackage{txfonts}
\usepackage{microtype}
\usepackage{ellipsis}
\usepackage{natbib}

\bibpunct{(}{)}{;}{a}{}{,}
\usepackage[pdftex]{hyperref}
\usepackage[xindy,nonumberlist,nopostdot,nogroupskip]{glossaries}

\makeglossaries
\usepackage[xindy]{imakeidx}
\makeindex

\begin{document} 
\title{Britannia Rule the Waves\thanks{Inspired by the patriotic hymn ``Rule, Britannia!'' written in the 18th century by T. A. Arne and J. Thompson.}}
\author{M. Nielbock \and T. Müller}
\institute{Haus der Astronomie, Campus MPIA, Königstuhl 17, D-69117 
Heidelberg, Germany\\
\email{nielbock@hda-hd.de}}

\date{Received July 27, 2016; accepted }

\abstract{The students are introduced to navigation in general and the longitude problem in particular. A few videos provide insight into scientific and historical facts related to the issue. Then, the students learn in two steps how longitude can be derived from time measurements. They first build a Longitude Clock that visualises the math behind the concept. They use it to determine the longitudes corresponding to five time measurements. In the second step, they assume the position of James Cook’s navigator and plot the location of seven destinations on Cook’s second voyage between 1772 and 1775.}

\keywords{Earth, navigation, countries, astronomy,  history, geography, Sun, equator, latitude, longitude, meridian,  celestial navigation, John Harrison, James Cook, clocks, exploration}

\maketitle
%

\section{Background information}

\subsection{Latitude and longitude}
\begin{figure}[!ht]
 \resizebox{\hsize}{!}{\includegraphics{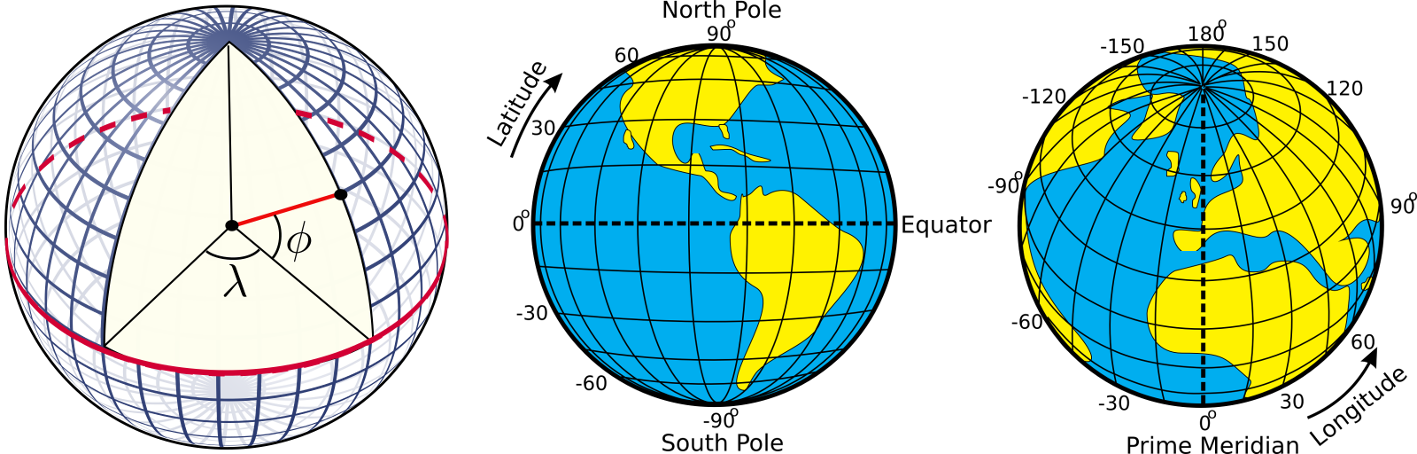}}
 \caption{Illustration of how the latitudes and longitudes of the Earth are 
defined (Peter Mercator, djexplo, CC0).}
 \label{f:latlong}
\end{figure}

Any location on an area is defined by two coordinates. The surface of a sphere 
is a curved area, but using coordinates like up and down does not make much 
sense, because the surface of a sphere has neither a beginning nor an ending. 
Instead, we can use
\newglossaryentry{spherical}
{
         name = {Spherical polar coordinates},
  description = {The natural coordinate system of a flat plane is Cartesian and       measures distances in two perpendicular directions (ahead, back, left, right). For a sphere, this is not very useful, because it has neither beginning nor ending. Instead, the fixed point is the centre of the sphere. When projected outside from the central position, any point on the surface of the sphere can be determined by two angles with one of them being related to the symmetry axis. Such axis defines two poles. In addition, there is the radius that represents the third dimension of space, which permits determining each point within a sphere. This defines the spherical polar coordinates. When defining points on the surface of a sphere, the radius stays constant.}
}
spherical polar coordinates originating from the centre of the sphere with the radius being fixed (Fig.~\ref{f:latlong}). Two angular coordinates remain. Applied to the Earth, they are called the latitude and the longitude. Its rotation provides the symmetry axis. The North Pole is defined as 
the point, where the theoretical axis of rotation meets the surface of the sphere and the rotation is counter-clockwise when looking at the North Pole from above. The opposite point is the South Pole. The equator is defined as the great circle
\newglossaryentry{great}
{
         name = {Great circle},
  description = {A circle on a sphere, whose radius is identical to the radius of the sphere.}
}
half way between the two poles.

The latitudes are circles parallel to the equator. They are counted from 
$0\degr$ at the equator to $\pm 90\degr$ at the poles. The longitudes are great 
circles connecting the two poles of the Earth. For a given position on Earth, 
the longitude going through the
\newglossaryentry{zenit}
{
         name = {Zenith},
  description = {Point in the sky directly above.}
}
zenith, the point directly above, is called the meridian. This is the line the Sun apparently
\newglossaryentry{appa}
{
         name = {Apparent movement},
  description = {Movement of celestial objects in the sky which in fact is caused by the rotation of the Earth.}
}
crosses at local noon. The origin of this coordinate is defined as the
\newglossaryentry{meri}
{
         name = {Meridian},
  description = {A line that connects North and South at the horizon via the zenith.}
}
Prime Meridian, and passes Greenwich, where the Royal Observatory of England is located. From there, longitudes are counted from $0\degr$ to $+180\degr$ (eastward) and $-180\degr$ (westward). 

Example: Heidelberg in Germany is located at 49\fdg4 North and 8\fdg7 East.

\subsection{Elevation of the pole (pole height)}
\begin{figure}[!ht]
 \resizebox{\hsize}{!}{\includegraphics{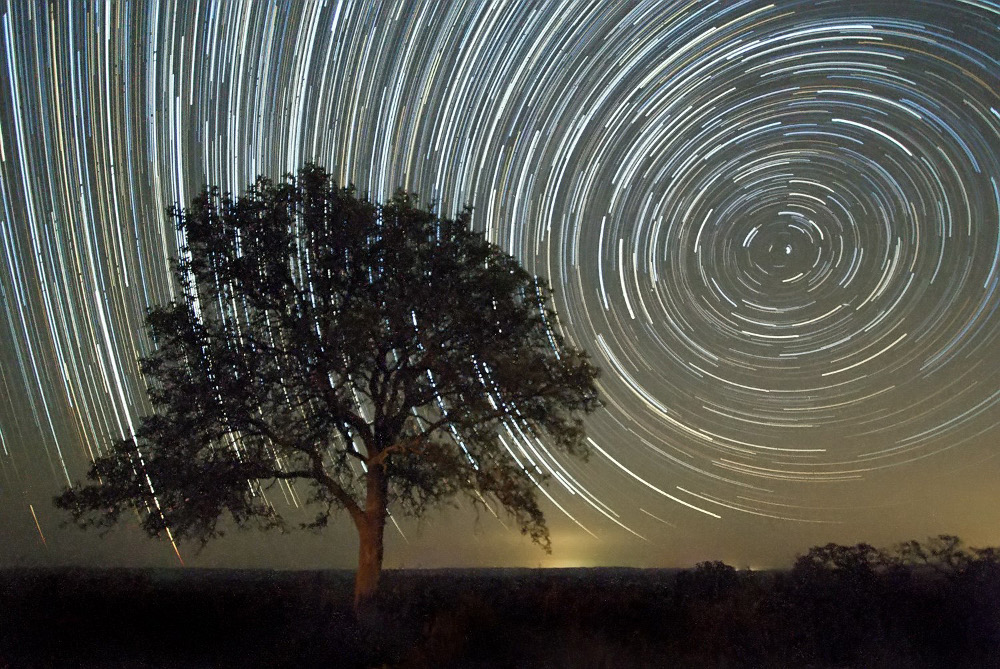}}
 \caption{Trails of stars at the sky after an exposure time of approximately 2 
hours (Ralph Arvesen, Live Oak star trails, 
\url{https://www.flickr.com/photos/rarvesen/9494908143}, 
\url{https://creativecommons.org/licenses/by/2.0/legalcode}).}
 \label{f:trails}
\end{figure}

If we project the terrestrial coordinate system of latitudes and longitudes at 
the sky, we get the celestial equatorial coordinate system. The Earth's equator 
becomes the celestial equator and the geographic poles are extrapolated to build 
the celestial poles. If we were to make a photograph with a long exposure of the 
northern sky, we would see from the trails of the stars that they all revolve 
about a common point, the northern celestial pole (Fig.~\ref{f:trails}).

\begin{figure}[!ht]
 \centering
 \resizebox{0.45\hsize}{!}{\includegraphics{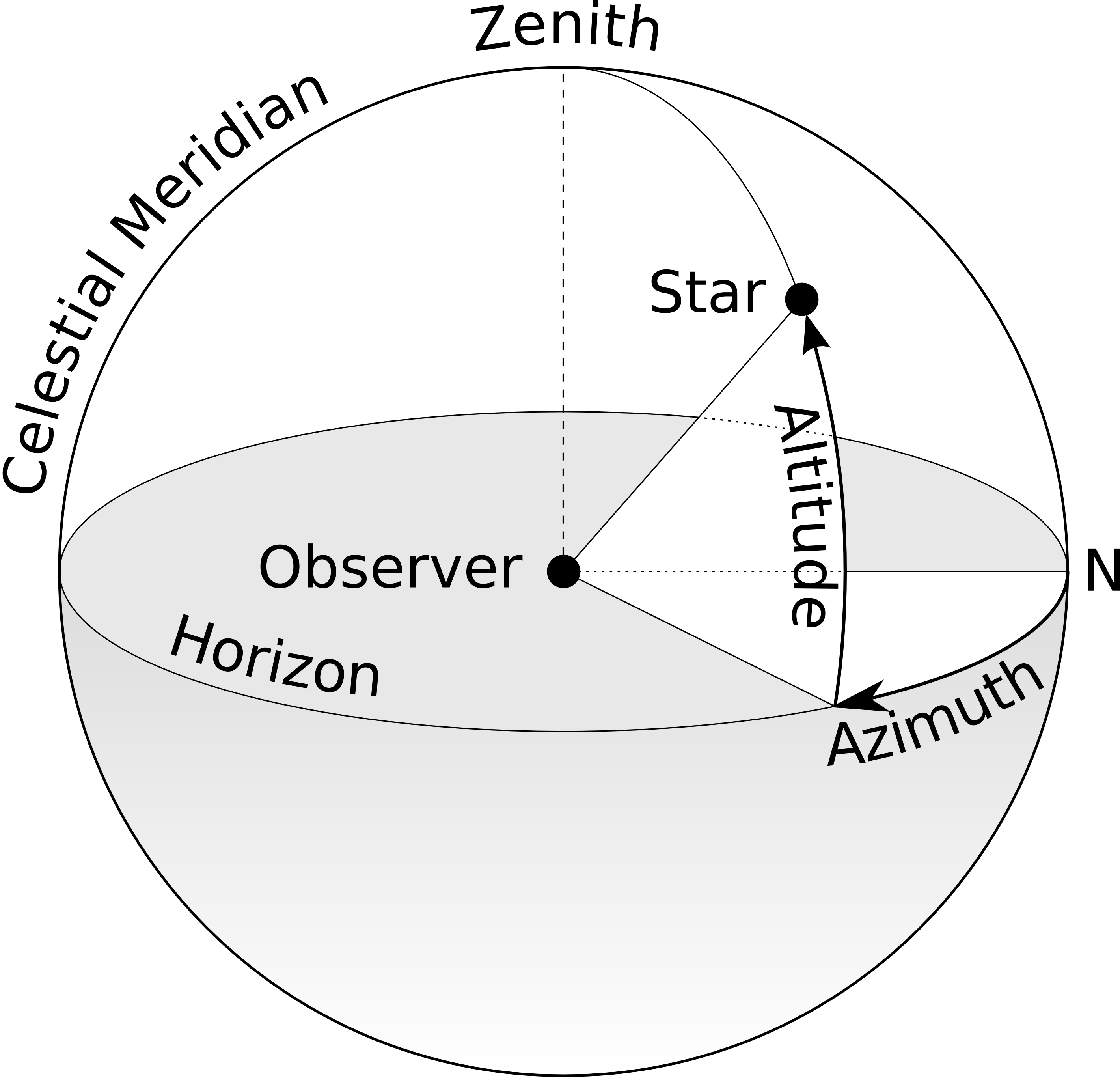}}
 \caption{Illustration of the horizontal coordinate system. The observer is the 
origin of the coordinates assigned as azimuth and altitude or elevation 
(TWCarlson, \url{
https://commons.wikimedia.org/wiki/File:Azimuth-Altitude_schematic.svg},
``Azimuth-Altitude 
schematic'', \url{https://creativecommons.org/licenses/by-sa/3.0/legalcode}).}
  \label{f:altaz}
\end{figure}

In the northern hemisphere, there is a moderately bright star near the celestial 
pole, the North Star or Polaris. At the southern celestial pole, there is no such star that can be observed with the naked eye. Other procedures have to be applied to find it. If we stood exactly at the geographic North Pole, Polaris would always be directly overhead. We can say that its elevation
\newglossaryentry{elev}
{
         name = {Elevation},
  description = {Angular distance between a celestial object and the horizon.}
}
would be (almost) $90\degr$. This information already introduces the horizontal coordinate system (Fig.~\ref{f:altaz}).

It is the natural reference we use every day. We, the observers, are the origin of that coordinate system located on a flat plane whose edge is the horizon. The sky is imagined as a hemisphere above. The angle between an object in the sky and the horizon is the altitude or elevation. The direction within the plane is given as an angle between $0\degr$ and $360\degr$, the azimuth, which is usually counted clockwise from north. In navigation, this is also called the bearing. The meridian is the line that connects North and South at the horizon and passes the zenith.

\begin{figure}[!ht]
 \resizebox{\hsize}{!}{\includegraphics{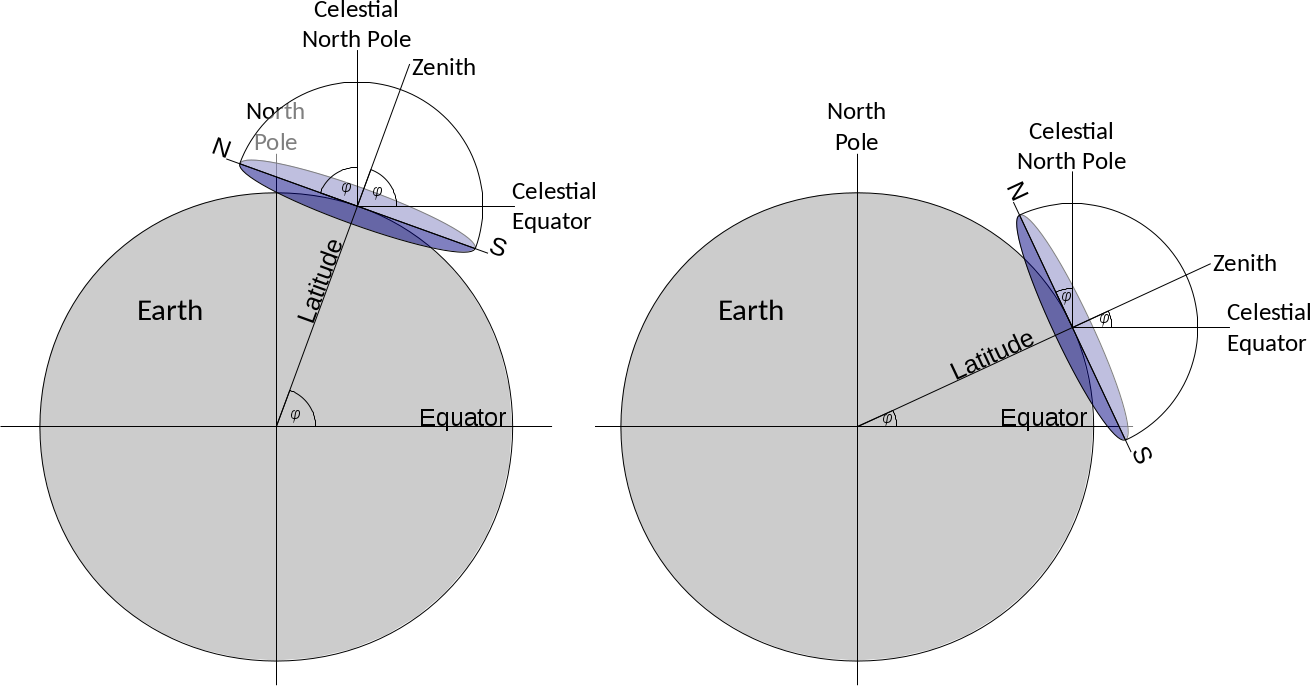}}
 \caption{When combining the three coordinate systems (terrestrial spherical, 
celestial equatorial, local horizontal), it becomes clear that the latitude of 
the observer is exactly the elevation of the celestial pole, also known as the 
pole height (own work).}
   \label{f:poleheight}
\end{figure}

For any other position on Earth, the celestial pole or Polaris would appear at 
an elevation smaller than $90\degr$. At the equator, it would just graze the 
horizon, i.e. be at an elevation of $0\degr$. The correlation between the 
latitude (North Pole = $90\degr$, Equator = $0\degr$) and the elevation of 
Polaris is no coincidence. Figure~\ref{f:poleheight} combines all three 
mentioned coordinate systems. For a given observer at any latitude on Earth, the 
local horizontal coordinate system touches the terrestrial spherical polar 
coordinate system at a single tangent point. The sketch demonstrates that the 
elevation of the celestial North Pole, also called the
\newglossaryentry{poleht}
{
         name = {Pole height},
  description = {Elevation of a celestial pole. Its value is identical to the latitude of the observer on Earth.}
}
pole height, is exactly the northern latitude of the observer on Earth.

\subsection{Local time and time zones}
\begin{figure}[!t]
 \resizebox{\hsize}{!}{\includegraphics{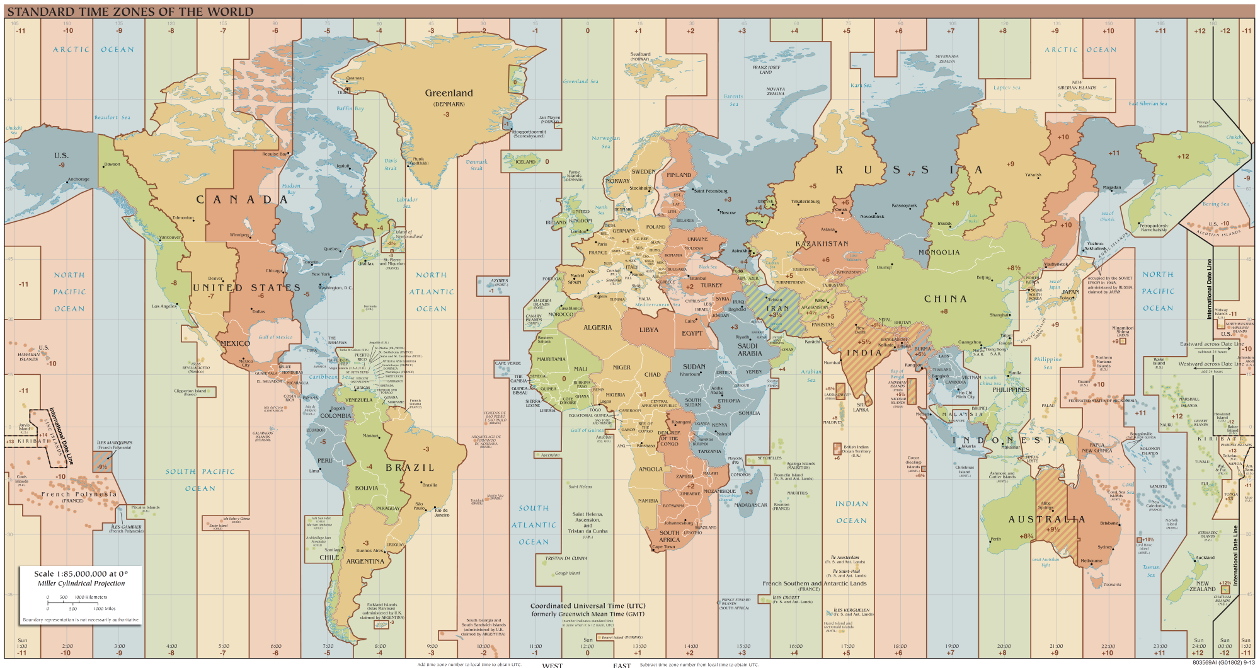}}
 \caption{World time zones. Instead of the local time that is based on the apparent path of the Sun in the sky and valid for single longitudes only, the common clocks show a time based on time zones which applies to many longitudes simultaneously (TimeZonesBoy, \url{https://commons.wikimedia.org/wiki/File:Standard_World_Time_Zones.png}, \url{https://creativecommons.org/licenses/by-sa/4.0/legalcode}).}
 \label{f:timezones}
\end{figure}

The Sun attains its highest elevation during the day when it crosses the local meridian. In the northern hemisphere, this is due south while in the southern hemisphere it is north. This is what defines local noon. Since the Earth rotates continuously, the apparent position of the Sun changes as well. This means that at any given point in time, local noon is actually defined for a single longitude only. However, clocks show a different time. Among other effects, this is mainly due to the time zones
\newglossaryentry{timez}
{
         name = {Time zone},
  description = {Before the advent of mass transportation across large distances by train, each town had its own local time that followed the solar time. This situation became impractical, as time tables for trains had to consider those shifts in time between stations. Therefore, in the 1840s it was decided to define a standard time that should be valid throughout Britain. Later, the concept was implemented all over the world with 24 zones of local standard times, the time zones. This is what we still use today.}
}
(Fig.~\ref{f:timezones}). Here, noon happens at many longitudes simultaneously. However, it is obvious that the Sun cannot transit the meridian for all those places at the same time. Therefore, the times provided by common clocks are detached from the ``natural'' local time a sundial shows.

\subsection{Mean and True Solar Time}
\begin{figure}[!ht]
 \centering
 \resizebox{0.4\hsize}{!}{\includegraphics{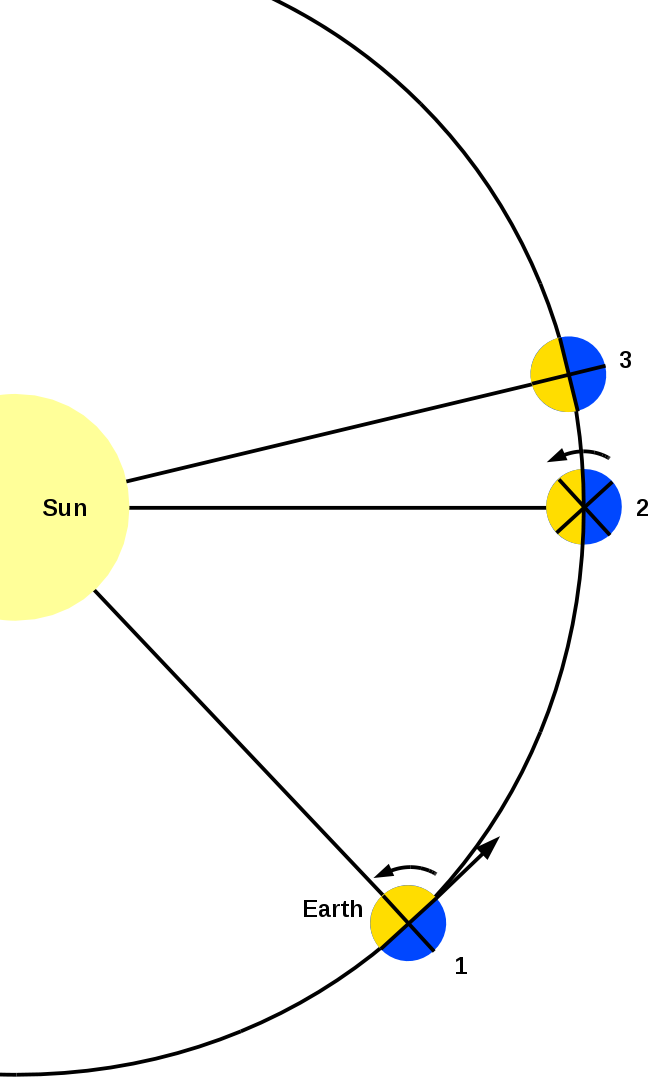}}
 \caption{Illustration of the difference between solar and sidereal day (own work).}
 \label{f:siderial}
\end{figure}

Figure~\ref{f:siderial} demonstrates the Earth’s rotation and its orbit around the Sun. The Sun illuminates the Earth, which leads to daytime (yellow) and night-time (blue). The cross that fills the Earth indicates its rotation angle. Within nearly 24 hours, the Earth rotates once around its own axis. As a result, the orientation relative to the sky at position 2 is the same as at position 1.

In addition to its own rotation, the Earth also revolves around the Sun. At position 1, the Sun is aligned with one spoke of the cross, i.e. it indicates local noon. However, at position 2, the Sun does not align with that same spoke anymore, i.e. it is at a different position in the sky. In order to have the Sun at the same spot in the sky again (next local noon), the Earth has to revolve and rotate for a little longer (position 3). As a result, a solar day lasts a few minutes longer than it takes to rotate around its own axis. The solar day takes almost exactly 24 hours.

However, the orbital speed of the Earth around the Sun is not constant throughout the year. It is faster near perihelion
\newglossaryentry{perih}
{
         name = {Perihelion},
  description = {This is the point of the Earth orbit that is closest to the Sun.}
}
and slower near
\newglossaryentry{aph}
{
         name = {Aphelion},
  description = {This is the point of the Earth orbit that is most distant to the Sun.}
}
aphelion.

Consequently, the duration of a true solar day changes constantly. This is reflected in the Apparent Solar Time (AST) or True Solar Time (TST)
\newglossaryentry{TST}
{
         name = {True Solar Time (Apparent Solar Time)},
  description = {The duration of a true solar day -- the period between two meridian passages of the Sun -- changes throughout the year. This is caused by the eccentric orbit of the Earth around the Sun. While the rotational speed of the Earth itself remains constant, the orbital velocity around the Sun is not. Consequently, a true solar day can be off from the mean value of 24 hours between about 20 seconds short and 30 seconds long. This accumulates to differences between the True and Mean Solar Time of up to approximately 15 minutes in either direction. In this time frame, noon is when the Sun is exactly on the meridian, i.e. south in the northern hemisphere and north in the southern hemisphere.}
}
which corresponds to the true apparent path of the Sun across the sky. Therefore, 12:00 noon TST is exactly, when the Sun is due south.

On average, the solar day lasts 24 hours, which corresponds to a full apparent revolution of the Sun in the sky, i.e. $360\degr$. We can now assume that this is true for each day of a year, which constitutes the
\newglossaryentry{MST}
{
         name = {Mean Solar Time},
  description = {The annual average of the duration of the Sun reaching the same azimuthal direction (e.g. between noons) is almost exactly 24 hours. The time measured according to those astronomical events is called the Mean Solar Time. In general, this differs from the time displayed by contemporary common clocks.}
}
Mean Solar Time (MST). This means that the angular speed $\omega$ of the Earth relative to the Sun’s apparent position for a solar day is on average $360\degr$ divided by 24 hours (h), or 15 degrees per hour:
\begin{equation}
 \omega = \frac{360\degr}{24\,{\rm h}}=15\frac{\degr}{\rm h}
\end{equation}
\subsection{Determining longitude}
With this rotational rate, one can determine longitude, if both the time at the Prime Meridian and the local time are known. If one calculates the difference between those times, the longitude is derived by simply multiplying this number with 15.

This concept was already proposed by the ancient Greek mathematician Hipparchus, who lived in the 2nd century BCE \citep{violatti_hipparchus_2013}.

Several methods have been tried and used in history to determine this time difference. Many involve the exact prediction of astronomical events that can be observed anywhere on Earth (eclipses, lunar distances to known bright stars, constellations of Galilean moons
\newglossaryentry{gali}
{
         name = {Galilean moon},
  description = {The four of more than 60 known moons of Jupiter (Io, Europa,
Callisto, Ganymede) that Galileo Galilei discovered in 1610 with one of the first astronomical telescopes used in human history.}
}
around Jupiter). Ships used to take along tables with the times at $0\degr$ longitude for such events. But they often turned out to be too hard to observe on a rocking ship.

The breakthrough was achieved by John Harrison, an 18th century clockmaker, who managed to invent highly accurate clocks that would even work on ships. His fourth version, the H4, had the design of a large pocket watch which always took along the local TST of Greenwich or, more precisely, of the Prime Meridian.

All the navigators had to do was to determine their local time, which was usually done at local noon, when the Sun passes the local meridian. The time difference in hours between local noon and TST is:
\begin{equation}
 \Delta t = 12\,{\rm h} - TST
\end{equation}
The longitude in degrees is then:
\begin{equation}
 \lambda = \Delta t \cdot 15\frac{\degr}{\rm h} = \left(12\,{\rm h} - TST\right)\cdot 15\frac{\degr}{\rm h}
\end{equation}
\begin{figure}[!t]
 \resizebox{\hsize}{!}{\includegraphics{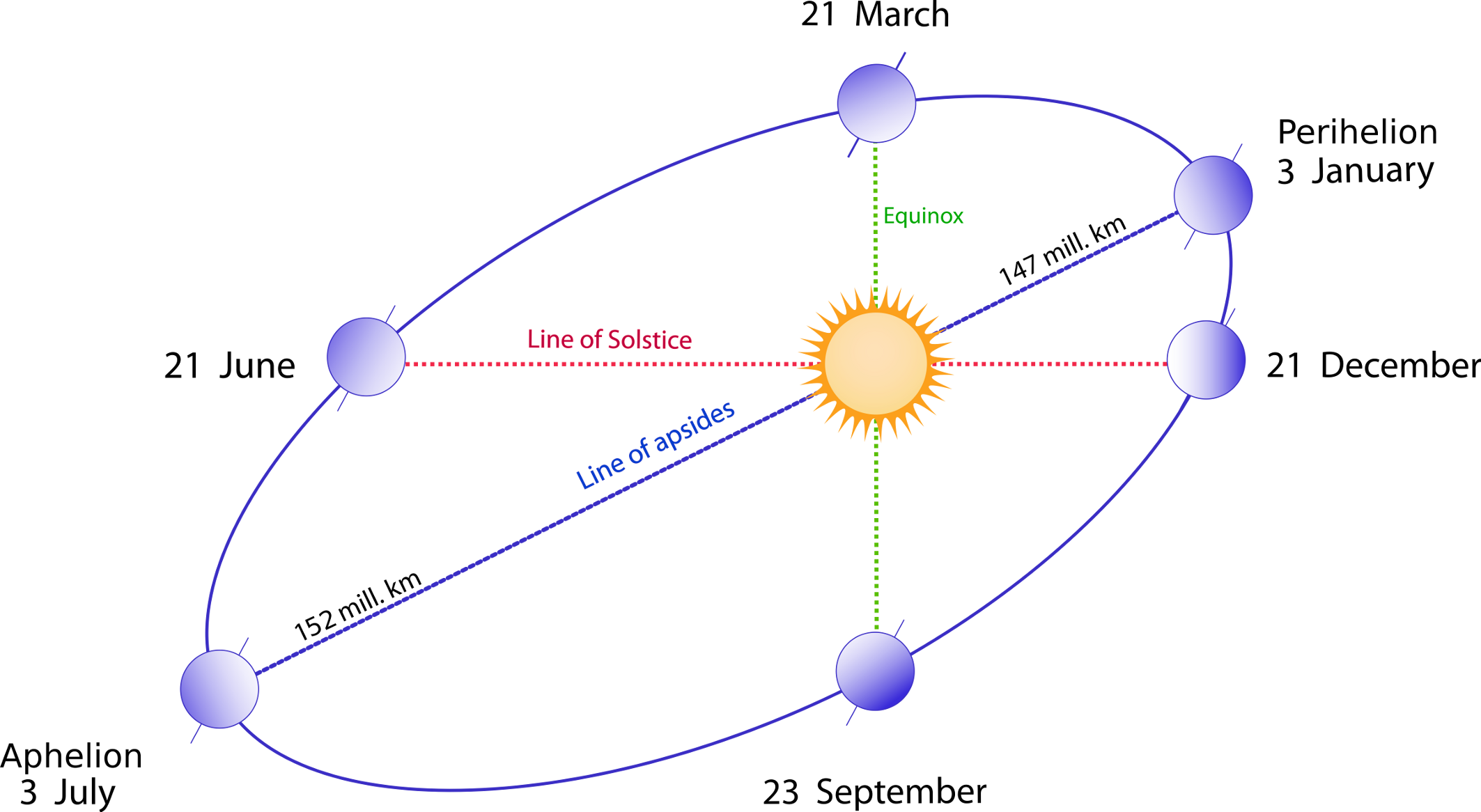}}
 \caption{Schematic view of the Earth's elliptical orbit around the Sun throughout a year. The position closest to the Sun is the perihelion, while the most distant point is the aphelion (following Duoduoduo's advice, vector image: Gothika, \url{https://commons.wikimedia.org/wiki/File:Seasons1.svg}, ``Seasons1'', annotations updated by Markus Nielbock, \url{https://creativecommons.org/licenses/by-sa/3.0/legalcode}).}
 \label{f:seasons}
\end{figure}

\subsection{The search for longitude}
While determining latitude with high accuracy has been possible for many centuries, tools and methods to determine longitude had been a long standing problem in navigation. Until the 18th century, navigators mostly had to rely on their experience. The only reasonably working method employed e.g. by early European explorers like Christoph Columbus was dead reckoning. This method is used to plot a ship's course by regularly recording its heading and speed. The tools employed for this were the magnetic compass and the log. The latter is a simple wooden board that is attached to a long rope wound on a reel. It had knots tied in regular distances. When thrown overboard, the log unrolls the rope. Counting the knots for a defined amount of time yields the ship's speed in knots (nautical miles per hour).

Unfortunately, there are several influences on open sea (wind, currents) that modify the course and speed. And such modifications are hard to estimate which often led to misjudgements and, not seldom, to catastrophic events.

\begin{figure}[!ht]
 \resizebox{\hsize}{!}{\includegraphics{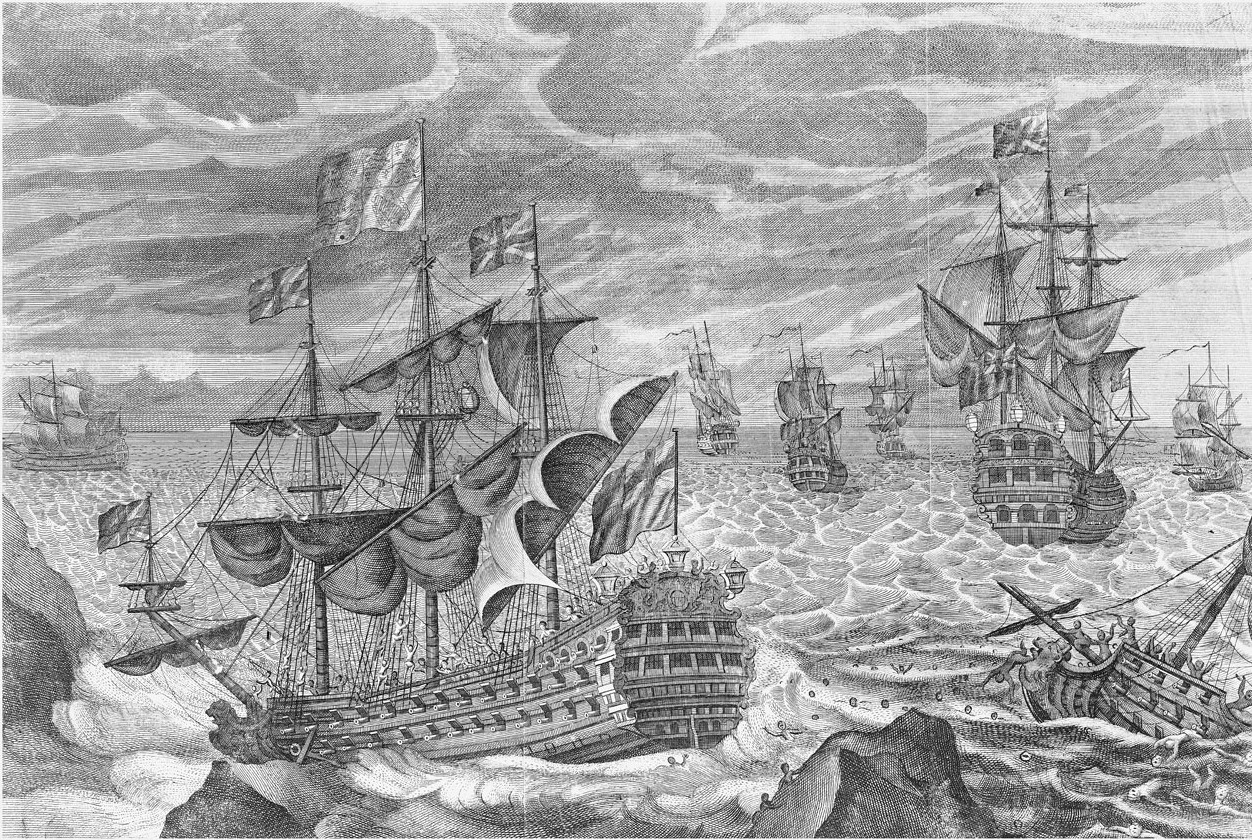}}
 \caption{Engraving from the 18th century showing the sinking HMS Association during the Scilly Islands naval disaster (\url{https://commons.wikimedia.org/wiki/File:HMS_Association_(1697).jpg}, public domain).}
 \label{f:scilly}
\end{figure}

\begin{figure*}[!t]
 \resizebox{\hsize}{!}{\includegraphics{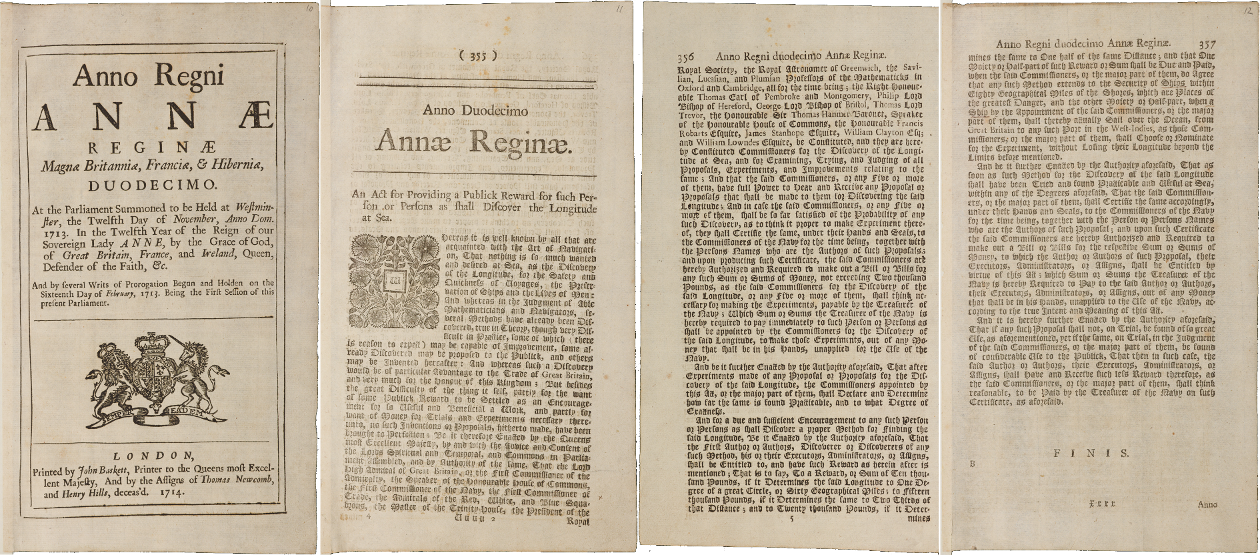}}
 \caption{Transcript of the initial version of the Longitude Act issued by the British Parliament in 1714 (Cambridge University Library, \url{https://cudl.lib.cam.ac.uk/view/MS-RGO-00014-00001/19}, \url{https://creativecommons.org/licenses/by-nc/3.0/legalcode}).}
 \label{f:lact}
\end{figure*}

One prominent example was the loss of a British fleet at Scilly Islands in 1707. On 22 October 1707, the navigators on board the flagship of the Commander-in-Chief of the British Fleets, Sir Cloudesley Shovell, the HMS Association, believed they were just entering the English Channel near Brittany. However, the island they saw belonged in fact to the Scilly Islands just west of Cornwall \citep{sobel_langengrad_2013}. When they noticed their mistake, it was already too late. Four of five ships were lost, and with them the lives of some 1500 sailors. Legend has it that poor Sir Shovell, who barely survived this disaster and just made it to the shores of the islands, was struck dead by a woman for a valuable emerald ring he wore on his fingers \citep{pickwell_improbable_1973,sobel_langengrad_2013}.

This naval catastrophe was probably the incident that convinced the British government that a better way to determine longitude was needed. In 1714, the Longitude Act was passed by the parliament of the United Kingdom \citep{higgitt_introduction_2015,sobel_langengrad_2013}. It provided rewards of up to \textsterling~20,000 for finding a method that allowed navigators to determine longitude within half a degree. A Board of Longitude was installed to evaluate the submissions.

Astronomical methods already existed, but they were either not accurate enough or impractical at sea. But there was one thing they all had in common: The sailors had to be able to determine the difference in time, Apparent Solar Time or True Solar Time that is, between their own position and the Prime Meridian. From this, one was able to infer the difference in angle the Earth had rotated between local noons of the two longitudes. Such methods included lunar eclipses, lunar distances to known bright stars, and configurations of the Galilean moons around Jupiter. All these events were tabulated for Greenwich local time and could be correlated to local times when observed at sea.

The most promising method of those was the lunar distance. However, neither the exact orbit of the Moon nor the positions of the stars were known accurately enough for navigational purposes. As a result, several observatories in Europe were founded to improve on this situation.

A much simpler method would have been to take along a clock that always displayed the time of the Prime Meridian. However, clocks manufactured until the early 18th century were neither accurate enough nor fit for sea voyages. This all changed with one person: John Harrison.

\subsection{John Harrison}
\begin{figure}[!ht]
 \centering
 \resizebox{0.6\hsize}{!}{\includegraphics{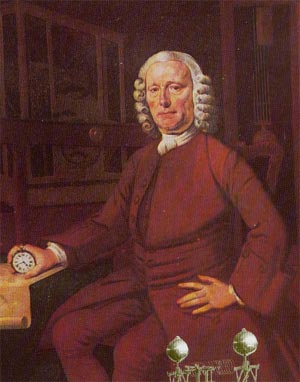}}
 \caption{Portrait of John Harrison (Oil painting by Thomas King, 1767, Science Museum, London, public domain).}
 \label{f:harrison}
\end{figure}

John Harrison (Fig.~\ref{f:harrison}) was an extraordinarily skilled English clockmaker of the 18th century. He made many inventions \citep{taylor_john_2007} that paved the road to the nautical chronometers which revolutionised navigation \citep{royal_museums_greenwich_longitude_2015,sobel_langengrad_2013}.

After building several pendulum and church tower clocks that reached an unprecedented precision and longevity with only little maintenance needed \citep{mcarthur-christie_240_2015}, he presented his first marine timekeeper in 1735, the \href{http://collections.rmg.co.uk/collections/objects/79139.html}{H1} \citep{betts_time_2006,sobel_langengrad_2013}. It was successfully tested during a journey to Lisbon and back. Harrison received several grants from the Board of Longitude to continue his work and to improve on this model. In 1759, he managed to present a revolutionary design of a compact watch, the \href{http://collections.rmg.co.uk/collections/objects/79142.html}{H4} \citep{shepherd_our_2013}. His son, William, took it on a transatlantic journey to Jamaica in 1761 which demonstrated its outstanding performance. The clock had only lost five seconds after being at sea for 81 days \citep{sobel_langengrad_2013}.

\subsection{Captain James Cook}
\begin{figure}[!t]
 \centering
 \resizebox{0.6\hsize}{!}{\includegraphics{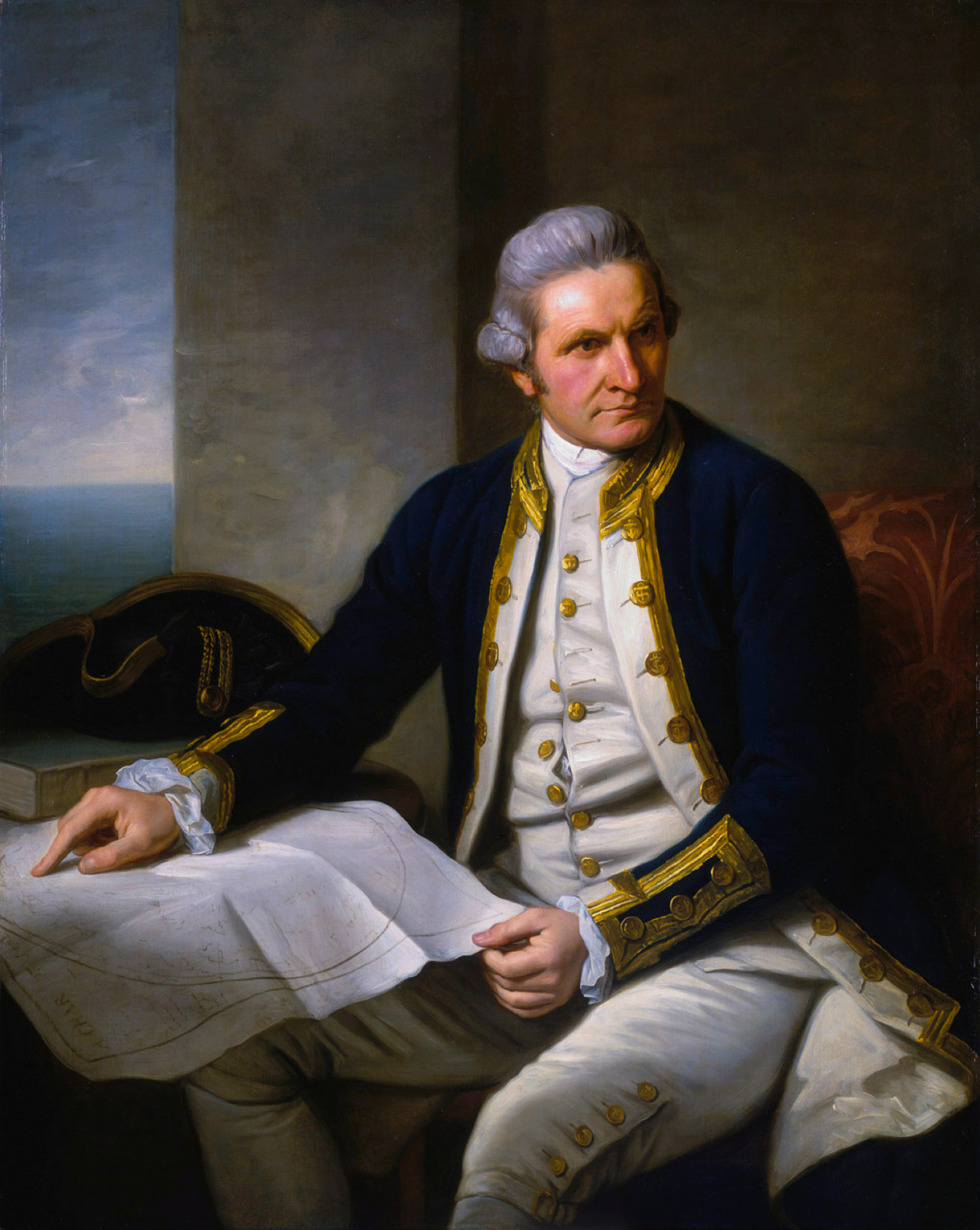}}
 \caption{Portrait of Captain James Cook (Painting by Nathaniel Dance-Holland, 1775-1776, National Maritime Museum, UK, public domain).}
 \label{f:cook}
\end{figure}

Captain James Cook (Fig.~\ref{f:cook}) was a British explorer, navigator and cartographer of the 18th century and a captain of the Royal Navy. He is famous for his three voyages to and through the Pacific Ocean. On his first voyage, Cook was the first to map the entire coastline on New Zealand and the eastern coast of Australia. He also made first contact with aboriginal tribes there. The spot of his first landfall was later named Botany Bay, just south of present-day Sydney \citep{cook_captain_2014}.

\begin{figure}[!t]
 \resizebox{\hsize}{!}{\includegraphics{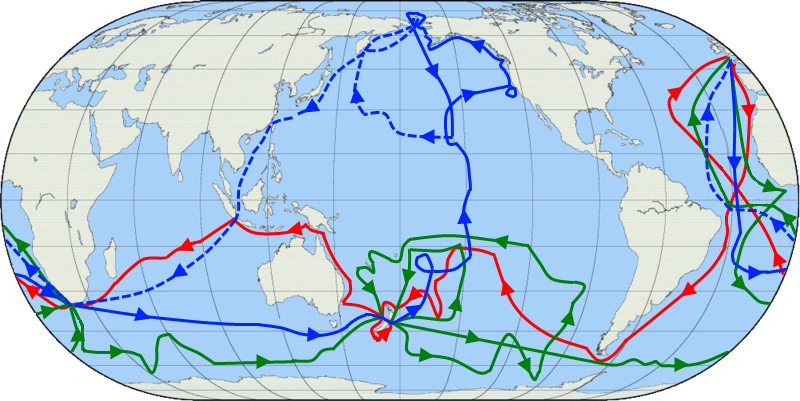}}
 \caption{Map showing the three voyages of Captain James Cook, with the first coloured in red, second in green, and third in blue. The route of Cook's crew following his death is shown as a dashed blue line (Jon Platek. Blank map by en:User:Reisio. \url{https://commons.wikimedia.org/wiki/File:Cook_Three_Voyages_59.png}, ``Cook Three Voyages 59'', \url{https://creativecommons.org/licenses/by-sa/3.0/legalcode}).}
 \label{f:cookvoyages}
\end{figure}

However, for our purposes, it is Cook’s second voyage from 1772 to 1775 \citep{cook_journal_1772} that interests us more (Fig.\ref{f:cookvoyages}). He took along a replica of John Harrison’s H4 watch to test its accuracy and its ability to determine longitude. It was manufactured in 1769 by Larcum Kendall and was known as the \href{http://collections.rmg.co.uk/collections/objects/79143.html}{K1} \citep{betts_time_2006}. It proved very reliable and contributed to the success of clocks for determining longitude. This method surely has also played a role in the success of the Global British Empire, which was mainly based on the ability of controlling the oceans and intercontinental trade.

\section{List of material}
The list contains items needed by one student.

\begin{itemize}
\item Worksheets
\item Crafting template and instructions: Longitude Clock
\item Crafting knife
\item Scissors
\item Glue
\item Calculator
\item Pencil
\item Computer (for Longitude Clock app, JAVA runtime environment installed)
\item Computer/tablet/smartphone with internet connection (for online map service)
\end{itemize}

\section{Goals}
With this activity, the students will learn that
\begin{itemize}
\item determining longitude reliably was an increasing problem in marine navigation of the 17th and 18th century  with a large number of ships lost at sea.
\item longitude can be derived from time measurements.
\item the astronomical (local) noon does not coincide with noon on the clock.
\item with John Harrison as a role model, persistence and conviction to a goal can lead to great achievements.
\end{itemize}

\section{Learning objectives}
The students  will be able to
\begin{itemize}
\item explain how time is related to the rotation of the Earth.
\item explain why determining the longitude on open sea had been difficult for centuries.
\item determine the longitude based on time measurements.
\item describe basic navigational skills used during the Age of Exploration.
\item name the English clockmaker who managed to build the first reliable marine chronometer.
\end{itemize}

\section{Target group details}

\noindent
Suggested age range: 14 -- 19 years\\
Suggested school level: Middle School, Secondary School\\
Duration: 90 minutes

\section{Evaluation}
When introducing the topic, ask the students, what defines a day. Using an Earth globe helps visualising the situation. After realising that it is the period of one full apparent revolution of the Sun around the Earth (e.g. passage of the meridian), let them work on activity 1.

Try to address methods of how to determine the time synchronously at two locations on the Earth. The time difference corresponds to the difference in longitude relative to a reference position.

Activity 2 is especially suited to understand how navigating with the Sun and a clock works. Guide the students through this task and check the results.

Let the students research the life and the achievements of John Harrison. There are suggested videos mentioned in the description of the activity.

\section{Full description of the activity}
\subsection{Preparations}
This activity only briefly introduces the concept of deriving latitude from angular measurements of celestial objects. A more detailed approach is provided by other activities of the same larger educational package focused on celestial navigation called ``Navigation Through the Ages''. The teacher might want to have a look at those first.

There are two versions of a Longitude Clock available. One is a computer app written in JAVA that keeps it independent of the operating system. However, please check, if it runs on the local computers. For details, see the instructions for the app below.

The other version is a hands-on cardboard dial, similar to a planisphere. The students will have to build them first. Instructions are included.

\subsection{Introduction}
It would be beneficial, if the activity be included into a larger context of seafaring, e.g. in geography, history, literature, etc.

Tip: This activity could be combined with other forms of acquiring knowledge like giving oral presentations in history, literature or geography highlighting navigation. This would prepare the field in a much more interactive way than what a teacher can achieve by summarising the facts. This topic is also well suited for acting classes.

Tip: There are good documentaries available highlighting the works of John Harrison and the history of finding longitude.

This is a suggested collection:

\medskip\noindent
``Longitude and latitude explained'', Australian National Maritime Museum (Duration: 2:33)\\
\url{https://www.youtube.com/watch?v=-8gg98ws2Eo}

\medskip\noindent
``Determine Longitude'', Science Online (Duration 11:10)\\
\url{https://www.youtube.com/watch?v=b7yoXhbOQ3Y}

\medskip\noindent
``The Clock That Changed the World (BBC History of the World)'', Leeds Museums (Duration 29:01)\\
\url{https://www.youtube.com/watch?v=T-g27KS0yiY}

\medskip\noindent
Let the students watch those during a preparatory lesson or at home.

\subsection{Questions, Answers and Discussion}
Ask the students, if they had an idea for how long mankind already uses ships to cross oceans. One may point out the spread of the Homo sapiens to islands and isolated continents like Australia.

\medskip\noindent{\em Possible answers:}\\
We know for sure that ships have been used to cross large distances already since 3,000 BCE or earlier. However, the early settlers of Australia must have found a way to cross the Oceans around 50,000 BCE

\medskip
Ask them, what could have been the benefit to try to explore the seas. Perhaps, someone knows historic cultures or peoples that were famous sailors. The teacher can support this with a few examples of ancient seafaring peoples, e.g. from the Mediterranean and the art of navigation.

\medskip\noindent{\em Possible answers:}\\
Finding new resources and food, trade, spirit of exploration, curiosity.

\medskip
Ask the students, how they find the way to school every day. What supports their orientation to not get lost? As soon as reference points (buildings, traffic lights, bus stops, etc.) have been mentioned ask the students, how navigators were able to find their way on the seas. In early times, they used sailing directions in connection to landmarks that can be recognised. But for this, the ships would have to stay close to the coast. Lighthouses improved the situation. But what could be used as reference points at open sea? Probably the students will soon mention celestial objects like the Sun, the Moon and stars.

\medskip\noindent{\em Additional suggested questions and answers}\\\noindent
Q: After watching the documentary, what were the major obstacles for building a marine timekeeper?\\\noindent
A: They were too inaccurate and unreliable at sea. Main reasons for this were the rolling movements of the ships interfering with the pendulums, the large changes in temperature as well as the lubrication. 

\medskip\noindent
Q: What triggered the Longitude Act, a call for finding an accurate method to determine longitude?\\\noindent
A: The naval disaster of 1707 at the Scilly Islands.

\medskip\noindent
Q: How are time and the rotation of the Earth connected?\\\noindent
A: The solar time, as we use it, is connected to the apparent diurnal
\newglossaryentry{diurn}
{
         name = {Diurnal},
  description = {Concerning a period that is caused by the daily rotation of the Earth around its axis.}
}
movement of the Sun. Every two noons are separated by 24 hours in which the Earth rotates (approximately) once around its own axis. The rotation of the Earth apparently moves the Sun around the sky. The longitude above which the Sun shines changes in time.

\medskip\noindent
Q: How long is one day in hours? How many degrees of one rotation are within one hour?\\\noindent
A: 1 day = 24 hours; 360 degrees = 1 full rotation; 15 degrees per hour

\medskip\noindent
Q: How would measuring time permit determining longitude?\\\noindent
A: The difference in time between the current position on Earth and a longitude reference (the Prime Meridian at Greenwich) is directly proportional to the longitude of the unknown position.

\medskip\noindent
Q: Who solved the longitude problem with a novel clock?\\\noindent
A: Clockmaker John Harrison

\medskip\noindent
Q: What was the clockwork of John Harrison's H1 clock made of?\\\noindent
A: wood

\medskip\noindent
Q: What is the main difference in design between H1 and H4?\\\noindent
A: The H1 is a large and heavy clock, while the H4 is similar to the pocket watch which is easier to operate.

\medskip\noindent
Q: Where are those clocks displayed now?\\\noindent
A: Greenwich Observatory Museums

\medskip\noindent
Q: Which great explorer tested and used a copy of the H4 during his voyages around the world?\\\noindent
A: James Cook

\medskip\noindent{\em Remark}\\\noindent
The following activity can either be done with the cardboard version of the Longitude Clock, or the computer app. Teachers are suggested to choose one. Please check, if the app runs on the local computers. For details, see the instructions for the app below.

\subsection{Building the Longitude Clock (available as separate document)}
Items needed:
\begin{itemize}
 \item Template printed on heavy paper or thin cardboard
 \item Instructions
 \item Crafting knife
 \item Scissors
 \item Glue
\end{itemize}

The template of the Longitude Clock consists of four pages.

\begin{figure}[!ht]
 \resizebox{\hsize}{!}{\includegraphics{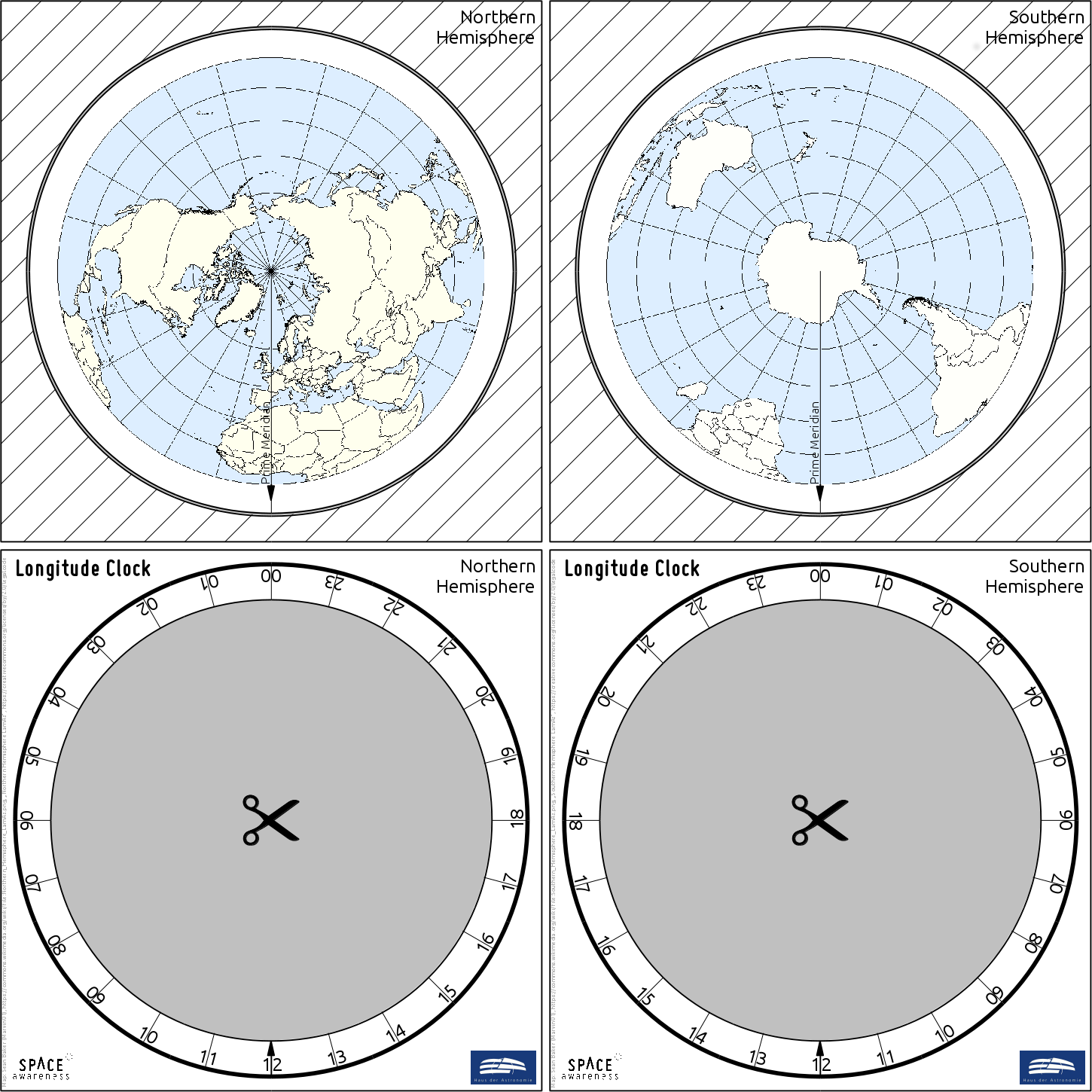}}
 \caption{Template for building the Longitude Clock. A printable version is available as a separate file (own work).}
 \label{f:template}
\end{figure}

\begin{enumerate}
 \item Print the template on extra heavy paper or cardboard to provide stability.
 \item Cut out the square areas.
 \item Glue the squares with the maps back to back. Make sure the glue is well distributed and the arrow on the Prime Meridian points to the same direction on both sides.
 \item Cut out the grey area inside the face of the clock (labelled “Longitude Clock”).
 \item After the glue has dried, cut off the hatched area around the maps, but do not destroy the hatched part. It is still needed later.
 \item Remove the grey area in between the hatched area and the maps. You may cut into the black borders surrounding it. Perhaps scissors help to trim the edges.
 \item Glue the part with the hatched area to the back of one of the faces of the Longitude Clock. Make sure the glue is well distributed on the hatched side. Let it dry.
 \item Put the disk with the maps inside the hatched area and check that it rotates smoothly. If needed, trim the edge some more. Then remove the disk again.
 \item Put glue on the remaining visible side of the hatched page.
 \item Carefully, put the disk with the maps inside. It must not receive any glue. Be sure that the correct side of the map disk is facing up. Double-check with the labelling of the clock face.
 \item Put on the back of the remaining face of the Longitude Clock on the glued hatched part.
 \item Let it dry and check that the disk rotates.
\end{enumerate}

\subsection{The Longitude Clock App}
There is a JAVA application attached to this unit that works in the same way as the Longitude Clock built by the students. After starting the software, the northern and southern hemispheres appear side by side. The time can be set by dragging with a computer mouse or entering the time. The software contains a readme file with further instructions.

\begin{figure}[!ht]
 \resizebox{\hsize}{!}{\includegraphics{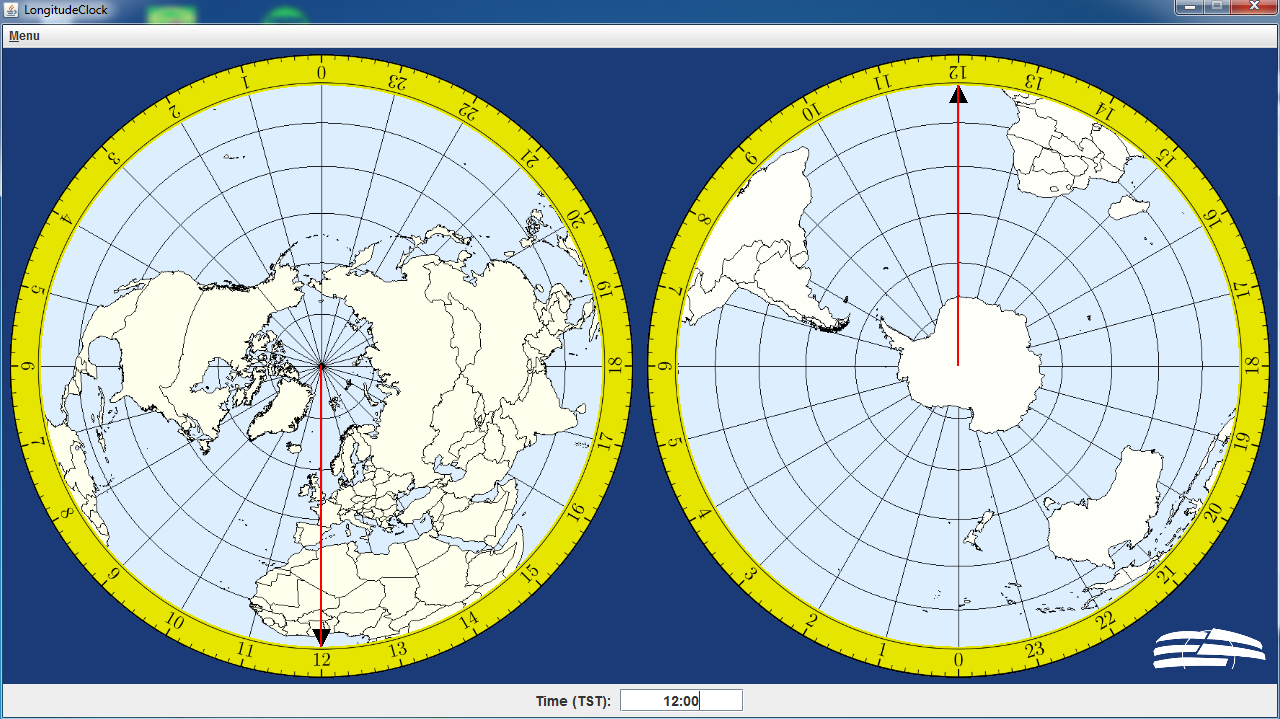}}
 \caption{Screen shot of the Longitude Clock application}
 \label{f:app}
\end{figure}

\subsubsection{Minimum requirements}
\begin{itemize}
 \item Java version 7 or higher
 \item Graphic board that supports at least OpenGL 3.3. When starting the application, the OpenGL version that is currently supported will be shown in a separate console.
\end{itemize}

\subsubsection{Important Remark}
Since the graphical standard mentioned above (OpenGL 3.3) was only introduced in 2010, it is well possible that the app will not run on all computers, especially such that are older or only possess a simple graphic card. It is recommended to test the application beforehand.

\subsubsection{Instructions}
Unzip the file \texttt{BritanniaRuleTheWaves\_LongitudeClockApp \_EUSPACE-AWE\_Navigation.zip} anywhere on a computer that has either Windows or Linux installed. A new folder called \texttt{LongitudeClock} is created. Go to that folder and run either the Windows or the Linux launcher script. Detailed information about its usage is included.

\subsection{Activity 1: Find the longitude}
Materials needed:
\begin{itemize}
 \item Worksheet
 \item Longitude Clock or/and Longitude Clock Application
 \item Pencil
 \item Calculator
 \item Computer, if the Longitude Clock app is used
\end{itemize}

The students will learn the concept of determining the longitude using mathematical equations and the Longitude Clock. Its precision is good enough to illustrate the procedure by visualising the underlying mathematical concept. However, the time resolution is too small to determine the longitude with very high precision as needed for navigation.

The worksheets contain a summary of the most important concepts needed to understand and carry out this activity.

\subsubsection{Using the Longitude Clock}
When navigating by sextant
\newglossaryentry{sext}
{
         name = {Sextant},
  description = {Navigational tool invented during the 18th century used to measure the angular altitude of celestial objects to determine latitude.}
}
and clock, the local time on board a ship is compared to the time measured at the Prime Meridian. For this purpose, ships used to carry along a highly accurate clock that is set to the time at $0\degr$ longitude, i.e. the time at Greenwich Observatory. The measurements were usually made at local noon, i.e. when the Sun attains its highest elevation.

The Prime Meridian is indicated on the Longitude Clock. To determine longitude, simply turn the marker of the Prime Meridian to the time displayed by the clock, which is set to the time of $0\degr$ longitude. The local longitude is then indicated at the time marker of 12 o'clock (local noon). The longitudes are indicated in steps of $15\degr$ west and east of the Prime Meridian. The teacher may choose to project the Longitude Clock app and demonstrate its usage to the students.

Note that for our exercises, we assume the clocks are showing the True Solar Time, but we calculate with the mean duration of a solar day of 24 hours.

\subsubsection{Exercise}
The worksheet contains a table with five examples of time readings (time at the Prime Meridian) for local noon (TST). The students calculate the time difference and the resulting longitude by applying the equations below. The results are then cross-checked with the Longitude Clock (paper or app version).

If TST is the true solar time at Greenwich (Prime Meridian), the time difference in hours between local noon and TST is:
\begin{equation}
 \Delta t = 12\,{\rm h} - TST
\end{equation}
The longitude corresponds to the angle the Earth had rotated between noon at the Prime Meridian and the local noon. Since the mean solar day lasts 24 hours, one hour corresponds to $15\degr$ in longitude. The local longitude in degrees is then:
\begin{equation}
\lambda = \Delta t \cdot 15\frac{\degr}{\rm h} = \left(12\,{\rm h} - TST\right) \cdot 15 \frac{\degr}{\rm h} 
\end{equation}
Negative values indicate western longitudes while positive values represent eastern longitudes.

\begin{table}[!h]
\centering
\caption{List of Greenwich time for which the students are asked to calculate the longitudes, if local noon is assumed. The solutions (not provided to the students) are added in italic writing.}
\label{t:a1}
\begin{tabular}{ccc}
\hline\hline
True Solar Time & \raisebox{-6pt}{$\Delta t ({\rm  h})$} & \raisebox{-6pt}{$\lambda (\degr)$} \\[-5pt]
at Greenwich (hh:mm)  &                       &                   \\
\hline
08:00 & \em +4  &  \em 60 East \\
23:00 & \em -11 & \em 165 West \\
18:00 & \em -6  &  \em 90 West \\
00:00 & \em +12 & \em 180 West/East \\
14:30 & \em -2.5 & \em 37.5 West \\
\hline
\end{tabular}
\end{table}

If the students have difficulties applying the equations, the teacher may want to demonstrate it by walking through the first example.

Note: The teacher may want to change the focus of the exercise by starting with the Longitude Clock and using the calculations as a cross-check instead.

\subsection{Actvity 2: Captain Cook's second voyage}
Materials needed:
\begin{itemize}
\item Worksheet
\item Pencil
\item Calculator
\item Computer/tablet/smartphone with internet connection
\end{itemize}

Using the worksheet, the students follow up on Cook's second voyage. They determine latitude and longitude of seven locations during the three years journey as provided in Tab.~\ref{t:a2} and locate each position on an on-line map.

\medskip\noindent
Q: How many minutes and seconds are in one hour?\\
A: 1 hour = 60 minutes = 3600 seconds

\medskip
The latitude can be calculated from any celestial object observed. If the position on the sky is known, the angle between the horizon and that object, the elevation, leads to the latitude. Celestial objects have coordinates of their own. Important is here the angle towards the equator. This angle is called declination, which corresponds to the latitude on Earth. Only at the terrestrial poles, the equator aligns with the horizon.

The latitude $\varphi$ is calculated from the declination $\delta$ and the elevation $\eta$ using the following equation.
\begin{equation}
\varphi =\pm \left(90\degr - \eta\right) + \delta
\end{equation}
The plus sign in front of the bracket is chosen, if the Sun attains its highest elevation to the south. It is minus, if the Sun is to the north. The sign of $\varphi$ is positive for northern latitudes and negative for southern latitudes. Unfortunately, the Sun changes its declination all the time. However, it can be calculated. For the seven measurements, its value is added to the table.

\subsubsection{Story}
Captain James Cook began his second voyage on 13~July~1772. His fleet consisted of two ships, the HMS Resolution and the HMS Adventure, the latter commanded by Captain Tobias Furneaux. Before setting sails, Cook took the first set of measurements.

After stops in the Madeira and Cape Verde Islands, the expedition anchored on 30~October 1772 at their first major southern port. They navigated around the Cape of Good Hope and after manoeuvring the ships through pack ice, they reached the Antarctic Circle on 17~January~1773. Both ships rendezvoused on 17~May~1773. From here, they explored the Pacific, and on 15~August reached an island, where the first pacific islander ever to visit Europe embarked on the HMS Adventure.

\begin{table*}
\centering
\caption{List of navigational measurements made on Cook's flagship HMS Resolution at seven dates during his second voyage. The measurements were all obtained at local noon, i.e. at the highest elevation of the Sun at that day. The times were obtained from the K1 watch James Cook took with him.}
\label{t:a2}
\begin{tabular}{ccccc}
\hline\hline
Date & Solar & Sun & Solar & True Solar Time \\
     & declination $(\degr)$ & direction & elevation $(\degr)$ & (hh:mm:ss) \\
\hline
13 July 1772    &  21.7 & South & 61.3 & 12:16:24 \\
30 October 1772 & -14.1 & North & 70.2 & 10:46:24 \\
17 May 1773     &  19.3 & North & 29.7 & 00:22:48 \\
15 August 1773  &  14.0 & North & 58.5 & 02:01:36 \\
30 January 1774 & -18.6 & North & 37.4 & 19:07:36 \\
17 December 1774& -23.4 & North & 60.0 & 17:05:14 \\
30 July 1775    &  18.5 & South & 58.1 & 12:06:00 \\
\hline
\end{tabular}
\end{table*}

The Adventure returned to England early, while Cook with the Resolution continued to roam the seas. After several attempts to venture south of the Antarctic Circle, he reached the most southern point on 30~January~1774, where ice blocked the passage. Cook continued to explore the Pacific, but finally decided to steer a course home. Cook headed east and his crew sighted land on 17~December~1774. They spent Christmas in a bay that Cook later named Christmas Sound.

He continued exploring the South Atlantic and discovered South Georgia and the South Sandwich Islands. After a stopover in southern Africa, the ship returned home on 30~July~1775.

\subsubsection{Exercise}
For seven of the destinations mentioned in this little report the table above lists measurements, from which the students should determine the latitude and the longitude, and add them to the table with the results below.

For the longitudes, use the equations of Activity 1. The times listed in the table have to be converted into hours with decimals representing the minutes and the seconds.

If available, check a map in an atlas or via a map service on-line, where on Earth these positions are. In Google Maps, simply enter the latitude followed by the longitude, both separated by a comma.

\subsubsection{Example}
The first measurement is at Cook's home port. It is taken on 13~July~1772 at 12:16:24. So, it is 12 hours, 16 minutes, and 24 seconds. To get this as hours with decimals, simply add up the following numbers:

\medskip\noindent
12 hours\\
16/60 hours\\
24/3600 hours

\medskip\noindent
The sum is rounded: 12.2733 hours

\medskip\noindent
Following the equation farther above, you get (rounded to the first decimal):
\begin{equation}
\lambda = \left(12\,{\rm h} - 12.2733\,{\rm h}\right) \cdot 15\degr/{\rm h} = -4.1\degr
\end{equation}
Thus, the longitude is $-4.1\degr$ or $4.1\degr$ west. To get the latitude, calculate (northern hemisphere, i.e. Sun is south):
\begin{equation}
\varphi = \left(90\degr - \eta\right) + \delta = \left(90\degr - 61.3\degr\right) + 21.7\degr
\end{equation}
\begin{table*}
\centering
\caption{Table prepared for the students to fill in the solutions. The results (not provided to the students) are added in italic writing.}
\label{t:a2solution}
\begin{tabular}{cccc}
\hline\hline
Date & Latitude $(\degr)$ & Longitude $(\degr)$ & Location on map \\
\hline
13 July 1772     & \em 50.4 N & \em   4.1 W & \em Plymouth \\
30 October 1772  & \em 33.9 S & \em  18.4 E & \em Cape Hope/Table Bay \\
17 May 1773      & \em 41.0 S & \em 174.3 E & \em Queen Charlotte Sound (NZ) \\
15 August 1773   & \em 17.5 S & \em 149.6 E & \em Tahiti \\
30 January 1774  & \em 71.2 S & \em 106.9 W & \em Most southern point, close to Antarctica \\
17 December 1774 & \em 53.4 S & \em  76.3 W & \em West of Patagonia, Strait of Magellan \\
30 July 1775     & \em 50.4 N &  \em  1.5 W & \em English Channel, close to Isle of Wight \\
\hline
\end{tabular}
\end{table*}

\subsection{Closure}
The students are invited to discuss how accurate this method is. The discussion may be guided along these questions. The answers are suggestions of where the discussion may lead.

\medskip\noindent
Q: Which steps are needed to plot a ship’s position on the open sea?\\
A: Either course and speed or solar elevation at local noon and Greenwich time.

\medskip\noindent
Q: How does weather interfere with this?\\
A: Sun must be visible to determine latitude and the time of local noon. Winds and storms make measurements difficult.

\medskip\noindent
Q: What knowledge and skills are needed to navigate in the way you did?\\
A: simple math, meaning of latitude and longitude, measuring angles of celestial objects, etc.

\medskip\noindent
Q: What skills and knowledge are needed to navigate with GPS?\\
A: very little

\section{Connection to school curriculum}
This activity is part of the Space Awareness category ``Navigation Through The Ages'' and related to the curricula topics:
\begin{itemize}
\item Coordinate systems
\item Basic concepts, latitude, longitude
\item Celestial navigation
\item Instruments
\end{itemize}

\section{Conclusion}
This module highlights the navigational challenge to determine longitude from time measurements. Along with an insight in the historic events that led to the invention of the first maritime timekeeper, the students exercise the method in two consecutive steps while they follow the steps of James Cook's second voyage around the world. Besides basic math and understanding the underlying fundamental concept, they build their own Longitude Clock that visualises in a simple way how determining longitude via time measurement works.

\begin{acknowledgements}
This resource was developed in the framework of Space Awareness. Space Awareness 
is funded by the European Commission’s Horizon 2020 Programme under grant 
agreement no. 638653.
\end{acknowledgements}

\bibliographystyle{aa}
\bibliography{Navigation}

\glsaddall
\printglossaries

\begin{appendix}
This unit is part of a larger educational package called “Navigation Through the Ages” that introduces several historical and modern techniques used for navigation. An overview is provided via:

\href{http://www.space-awareness.org/media/activities/attach/b3cd8f59-6503-43b3-a9e4-440bf7abf70f/Navigation\%20through\%20the\%20ages\%20compl\_z6wSkvW.pdf}{Navigation\_through\_the\_Ages.pdf}

 \section{Supplemental material}
 The supplemental material is available on-line via the Space Awareness project website at \url{http://www.space-awareness.org}. The direct download links are listed as follows:
 
 \begin{itemize}
  \item Worksheets: \href{https://drive.google.com/file/d/0Bzo1-KZyHftXZGdhaFZsdWdhRkU/view?usp=sharing}{astroedu1646-Britannia-Rule-the-Waves-WS.pdf}
  \item Longitude Clock template: \href{https://drive.google.com/file/d/0Bzo1-KZyHftXdDdOaXN0ZDdVT2s/view?usp=sharing}{astroedu1646-Britannia-Rule-the-Waves-Template.pdf}
  \item Longitude Clock App: \href{https://drive.google.com/file/d/0Bzo1-KZyHftXdXVSZF9MMm5fZXM/view?usp=sharing}{astroedu1646-Britannia-Rule-the-Waves-App.zip}
\end{itemize}
\end{appendix}
\end{document}